\documentclass[twocolumn,aps,superscriptaddress,showpacs,prb]{revtex4}
\usepackage{graphicx}
\usepackage{times}
\usepackage{amsmath}
\usepackage{amssymb}
\usepackage{units}
\usepackage{stmaryrd} 

\usepackage{color}
\usepackage{dcolumn}
\usepackage{bm}

\usepackage{color}

\DeclareGraphicsExtensions{.pdf,.png,.eps ,.jpg}

\begin{document}
\preprint{0}

\title{Momentum resolved spin dynamics of bulk and surface excited states in the topological insulator $\mathrm{Bi_{2}Se_{3}}$}

\author{C. Cacho}
\affiliation{Central Laser Facility, STFC Rutherford Appleton Laboratory, Harwell, United Kingdom}

\author{A. Crepaldi}
\affiliation{Elettra - Sincrotrone Trieste S.C.p.A, Strada Statale 14, km 163.5, 34149 Basovizza, Trieste, Italy}

\author{M. Battiato}
\affiliation{Institute of Solid State Physics, Vienna University of Technology, Wiedner Hauptstrasse 8-10, A 1040 Wien, Austria}

\author{J. Braun}
\affiliation{Department Chemie, Ludwig-Maximilians-Universit\"at M\"unchen, Butenandtstrasse 5-13, 81377 M\"unchen, Germany}

\author{F. Cilento}
\affiliation{Elettra - Sincrotrone Trieste S.C.p.A, Strada Statale 14, km 163.5, 34149 Basovizza, Trieste, Italy}

\author{M. Zacchigna}
\affiliation{C.N.R. - I.O.M., Strada Statale 14 km 163.5 Trieste, Italy}

\author{M. C. Richter}
\affiliation{Laboratoire de Physique des MatŽriaux et des Surfaces, Universit\'e de Cergy-Pontoise, 5 mail Gay-Lussac, 95031 Cergy-Pontoise, France}
\affiliation{DSM, IRAMIS, Service de Physique de l'Etat Condens\'e, CEA-Saclay, 91191 Gif-sur-Yvette, France }

\author{O. Heckmann}
\affiliation{Laboratoire de Physique des MatŽriaux et des Surfaces, Universit\'e de Cergy-Pontoise, 5 mail Gay-Lussac, 95031 Cergy-Pontoise, France}
\affiliation{DSM, IRAMIS, Service de Physique de l'Etat Condens\'e, CEA-Saclay, 91191 Gif-sur-Yvette, France }

\author{E. Springate}
\affiliation{Central Laser Facility, STFC Rutherford Appleton Laboratory, Harwell, United Kingdom}

\author{Y. Liu}
\affiliation{Diamond Light Source, Chilton, Didcot, Oxfordshire, United Kingdom}

\author{S. S. Dhesi}
\affiliation{Diamond Light Source, Chilton, Didcot, Oxfordshire, United Kingdom}

\author{H. Berger}
\affiliation{Institute of Condensed Matter Physics (ICMP), \'Ecole Polytechnique F\'ed\'erale de Lausanne (EPFL), CH-1015 Lausanne, Switzerland}

\author{Ph. Bugnon}
\affiliation{Institute of Condensed Matter Physics (ICMP), \'Ecole Polytechnique F\'ed\'erale de Lausanne (EPFL), CH-1015 Lausanne, Switzerland}

\author{K. Held}
\affiliation{Institute of Solid State Physics, Vienna University of Technology, Wiedner Hauptstrasse 8-10, A 1040 Wien, Austria}

\author{M. Grioni}
\affiliation{Institute of Condensed Matter Physics (ICMP), \'Ecole Polytechnique F\'ed\'erale de Lausanne (EPFL), CH-1015 Lausanne, Switzerland}

\author{H. Ebert}
\affiliation{Department Chemie, Ludwig-Maximilians-Universit\"at M\"unchen, Butenandtstrasse 5-13, 81377 M\"unchen, Germany}

\author{K. Hricovini}
\affiliation{Laboratoire de Physique des MatŽriaux et des Surfaces, Universit\'e de Cergy-Pontoise, 5 mail Gay-Lussac, 95031 Cergy-Pontoise, France}
\affiliation{DSM, IRAMIS, Service de Physique de l'Etat Condens\'e, CEA-Saclay, 91191 Gif-sur-Yvette, France }

\author{J. Min\'ar}
\affiliation{Department Chemie, Ludwig-Maximilians-Universit\"at M\"unchen, Butenandtstrasse 5-13, 81377 M\"unchen, Germany}
\affiliation{New Technologies-Research Center, University of West Bohemia, Univerzitni 8, 306 14 Pilsen, Czech Republic}

\author{F. Parmigiani}
\email{fulvio.parmigiani@elettra.eu}
\affiliation{Elettra - Sincrotrone Trieste S.C.p.A, Strada Statale 14, km 163.5, 34149 Basovizza, Trieste, Italy}
\affiliation{Universit\`a degli Studi di Trieste - Via A. Valerio 2, Trieste 34127, Italy}
\affiliation{International Faculty - University of K\"oln, Germany}


\date{\today}


\begin{abstract}

The prospective of optically inducing a spin polarized current for spintronic devices has generated a vast interest in the out-of-equilibrium electronic and spin structure of topological insulators (TIs). In this Letter we prove that only by measuring the spin intensity signal over several order of magnitude in spin, time and angle resolved photoemission spectroscopy (STAR-PES) experiments is it possible to comprehensively describe the optically excited electronic states in TIs materials. The experiments performed on $\mathrm{Bi_{2}Se_{3}}$ reveal the existence of a Surface-Resonance-State in the 2nd bulk band gap interpreted on the basis of fully relativistic \emph{ab-initio} spin resolved photoemission calculations. Remarkably, the spin dependent relaxation of the hot carriers is well reproduced by a spin dynamics model considering two non-interacting electronic systems, derived from the excited surface and bulk states, with different electronic temperatures.

\end{abstract}


\maketitle


The possibility of optically inducing a spin-polarized electrical current in topological insulators (TIs)  \cite{Hasan_RMP_2010, Qi_RMP_2011, McIver_Nnano_2012} has recently bred an increasing interest in the out-of-equilibrium properties of these materials \cite{Sobota_PRL_2012, Perfetti_arxiv_2012, Gedik_PRL_2012, Crepaldi_2012, Hajlaoui_EPJ_2013, Crepaldi_PRB_2013, Wang_Science_2013, Marsi_natCom_2014, Sobota2014, Niesner_PRB_2014}. In this context, TR-ARPES experiments have provided important information on the electron-phonon coupling strength \cite{Gedik_PRL_2012, Crepaldi_2012, Crepaldi_PRB_2013}, the electron and hole diffusion \cite{Marsi_natCom_2014}  and the scattering between the topologically protected surface state (TSS) and the bulk conduction band (BCB)  \cite{Gedik_PRL_2012, Perfetti_arxiv_2012, Niesner_PRB_2014}. In parallel static spin-resolved studies have confirmed the helical spin structure of the TSS \cite{Hasan_RMP_2010, Hsieh_nature_2009, Hsieh_Science_2009}. However, despite the large difference in the spin properties between the bulk and the surface states, there is no direct information about the spin dynamics of these bands because of the lack of spin resolution in the TR-ARPES experiments so far reported in literature \cite{Sobota_PRL_2012, Perfetti_arxiv_2012, Gedik_PRL_2012, Crepaldi_2012, Hajlaoui_EPJ_2013, Crepaldi_PRB_2013, Wang_Science_2013, Marsi_natCom_2014, Sobota2014, Niesner_PRB_2014}.


\begin{figure*}
  \includegraphics[width = 1.0\textwidth]{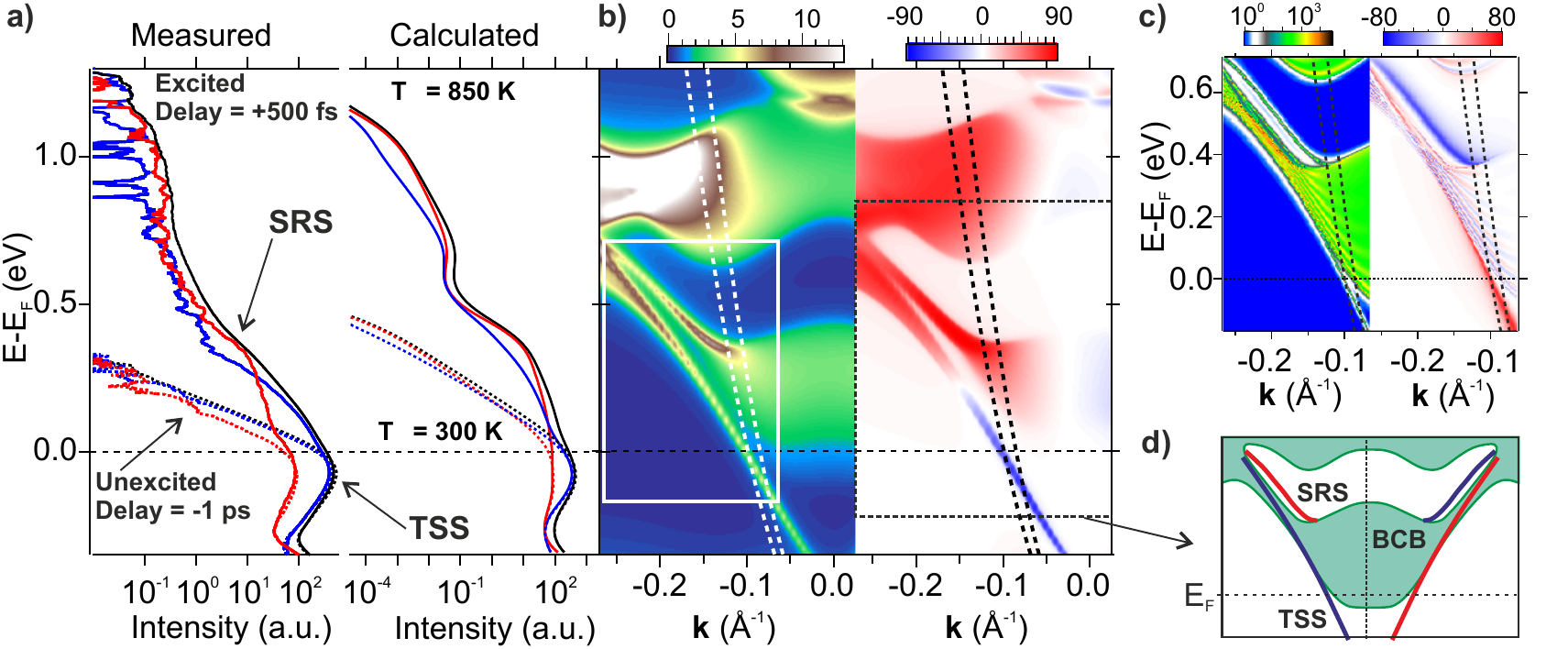}
  \caption[didascalia]{(Color online) (a) measured (left) and calculated (right) spin resolved EDCs along the $\Gamma K$ high symmetry direction close to the TSS $k_F$ of $\mathrm{Bi_{2}Se_{3}}$.  Dotted (continuous) blue and red lines indicate the two opposite spin intensities before (after) optical excitation, while black lines are the sum of the two spin signals.  The calculated EDCs are extracted from the photo-emission calculations of panel~(b), for T = 300 K (dotted line) and T = 850 K (continuous line), within the dashed lines. (b) \emph{ab initio} fully relativistic photoemission calculations for $\mathrm{Bi_{2}Se_{3}}$ along the $\Gamma K$ high symmetry direction with 6.2 eV photon energy and \emph{s} polarization. (c) calculated ground state spin polarized electronic properties in the white rectangular area of panel (b), in order to isolate the photo-emission matrix element effects. (d) schematics of the electronic properties of $\mathrm{Bi_{2}Se_{3}}$, in particular the TSS and SRS states dispersing across the band gap within the BCB.
   }
  \label{fig:arpes1}
\end{figure*}


In this letter we report on a combined experimental and theoretical spin- and time-resolved ARPES study (STAR-PES) of the archetypal TI, $\mathrm{Bi_{2}Se_{3}}$, optically excited by infrared (IR) ultrafast ($\sim150$fs) laser pulses. Our work reveals the presence of a Surface-Resonance-State (SRS) localized in the projected bulk band gap in very good agreement with fully relativistic \emph{ab-initio}  photoemission calculations. Furthermore, the photo-induced hot carrier exhibit very different energy dependent dynamics for both  spin channels. Remarkably, the measured spin relaxation is well accounted by considering two distinct electronic populations in the surface and bulk states, thermalized with two different electronic temperatures and cooling times. The out-of-equilibrium properties of the two electronic populations are well approximated by two non-interacting electronic systems, where neither energy nor particles are exchanged in the investigated time window ($\sim8$~ps). 



Figure 1 (a) shows on a logarithmic scale the measured and calculated spin resolved energy distribution curves (EDCs). The experimental data have been acquired in proximity of the TSS Fermi wave-vector,  at $-7^{\circ}$ \emph{i.e.} between the two dashed lines of Fig.~1~(b) along the  $\Gamma K$ high symmetry direction $k_F$, using a 6.2 eV photon energy and \emph{s} polarization.  The full equilibrium spin resolved and the spin integrated time-resolved ARPES  are shown in the supplementary material, along with the description of the experimental setup \cite{Suppl_mat}. Blue and red correspond to the two opposite spin directions, dotted (continuous) line indicates the measured EDCs before (after) the arrival of the 1.55 eV pump pulse at the delays of -1 ps and +500 fs, respectively. Hereafter we use red and blue to indicate spin up and down, respectively.
 We associate the spin polarized peak, located slightly below the Fermi level $E_F$, to the TSS approaching $k_F$. Before optical excitation, both spin-EDCs decrease exponentially above $E_F$, showing a characteristic Fermi Dirac cut-off.  When the system is optically excited, the spin EDCs reveal more complex features.  In particular, at $\sim$ 250 meV above $E_F $ the spin polarization is inverted with respect to TSS, while between 500 meV and 800 meV, a spin unpolarized region is observed. At $E -E_F > 1$ eV the spin component  (red), which had minority character in the TSS, becomes dominant, within the experimental noise. In agreement with the helical spin texture of the TSS \cite{Hasan_RMP_2010, Hsieh_nature_2009, Jozwiak_NatPhys_2013, Zhu_PRL_2014},  similar behavior, but with opposite spin, is  observed at opposite $k_F$ (see supplementary material \cite{Suppl_mat}). 



To reveal the origin of the photo-excited spin structure we performed \emph{ab-initio} spin-resolved photoemission calculations for 6.2 eV photon energy and s-polarized light, corresponding to the present experimental conditions. These spin ARPES intensity calculations are based on the relativistic \emph{one-step} model. The model, in its spin-density matrix formulation, describes properly the complete spin-polarization, \emph{i.e.} all three components of the spin-polarization vector for each ($ k_{x}, k_{y}$) point \cite{Suppl_mat}. The final state is modeled as a so called time-reversed spin-polarized low-energy electron diffraction (SPLEED) state \cite{Bra96}. The photoemission calculations accounts also for matrix-element effects and multiple scattering effects in the initial and final states. Many-body effects are included phenomenologically in the SPLEED calculations, by using a parameterized and weakly energy-dependent complex inner potential, $V_0(E)=V_{0{\rm r}} (E)+iV_{0{\rm i}}(E)$ \cite{Pen74}. 


Figure 1 (a) right panel shows the calculated spin-EDCs extracted from k-dependent photoemission calculations multiplied by a Fermi-Dirac distribution function. Dotted (continuous) blue and red lines correspond to an electronic temperature equal to $T_{(-1 ps)} = 300$ K ($T_{(500 fs)} = 850$ K) where 850 K is the optimal temperature to reproduce the measured spin integrated intensity (black line) at +500 fs delay. This procedure constitutes a benchmark for our experimental data, in order to understand whether a single electronic distribution is sufficient to reproduce our data.

Figure 1 (b) shows the spin integrated (left) and spin polarization (right) calculated photoemission intensity along the $\Gamma K$ high symmetry direction. Due to the influence of the light polarization on the photoelectron spin, the ground state calculations are also required to completely interpret the photoemission spin polarization  \cite{ Jozwiak_NatPhys_2013, Zhu_PRL_2014,Minar_PRX_2014}. Figure 1 (c) presents the ground state calculations over a selected region (see Figure 1 (b) white rectangle). The dotted curves indicate the region where the EDCs of panel (a) are integrated. Beside the TSS and the BCB, an additional spin polarized surface state, dispersing in a bulk projected band gap, is observed in the experimental data and consistently reproduced by the calculated spin-resolved spectra. The surface character of this state, never reported before, differs from the TSS, and it is identified as a surface resonance of bulk states with a large bulk contribution. Noticeably, the SRS and TSS  have opposite spin polarization, and the former is topologically \emph{trivial}. Figure 1 (d) schematizes the photoelectron spin properties of $\mathrm{Bi_{2}Se_{3}}$, with the spin polarized TSS and SRS dispersing in two different projected band gaps of the BCB.

The photoemission spin structure reported  in Fig.~1~(b) shows that for \emph{s}-polarized light the TSS and SRS photoelectron spin is opposite to the ground state polarization of Fig.~1~(c). This is a consequence of the strong spin-orbit coupling combined with the  orbital-dependent photo-excitation probabilities \cite{Minar_PRX_2014}. We notice that the electron spin polarization in the ground state of the TSS and the SRS is opposite. Furthermore, even though the BCB ground state is spin unpolarized, the BCB reveals a photoelectron spin polarization, increasing from zero at $E_F$ to appreciable polarization at $E- E_F \sim0.9$~eV. This is  a consequence of the partial hybridization between the BCB and the TSS states. 


 \begin{figure}
  \includegraphics[width = 0.5\textwidth]{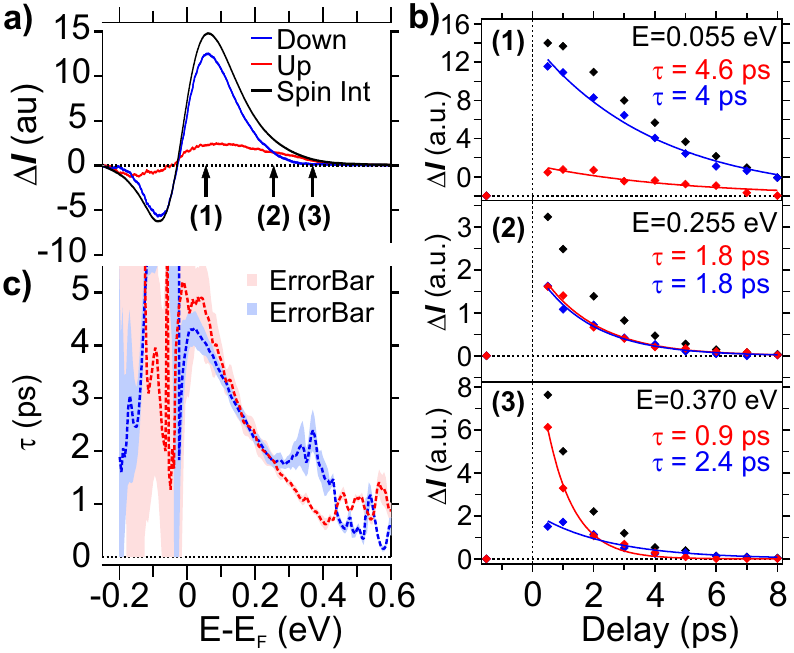}
  \caption[didascalia]{(Color online) (a) Delta spin EDCs obtained as the difference between the signal measured after and before optical excitation are shown. Arrows point towards the energies where the signal is integrated for extracting the energy dependent spin-dynamics shown in panel~(b). (b) time evolution of the intensity measured for spin up (red) and down (blue),  while black markers indicate the total intensity sum of the two spin components. From these the spin dependent relaxation times are extracted from these traces via an exponential fit. The characteristic relaxation times are shown in panel (c) for the spin up (red) and spin down (blue) as a function of the binding energy. 
   }
  \label{fig:arpes2}
\end{figure}



The measured spin inversion at  $E -E_F \sim250$~meV, as observed in Fig.1 (a), is assigned to the SRS,  having a spin polarization opposite to the TSS. The polarization detected at high energy ($E -E_F >1$~eV) results from the photoemission from the top-side of the BCB, whose weak spin polarization is visible in Fig. 1 (b). Finally, the spin unpolarized weak signal between  500 meV and 800 meV corresponds to the energy region of  the projected band gap.


At first glance, the comparison between the experimental and calculated spin EDCs in Fig.~1~(a ) might suggest the possibility of describing  the optically excited EDCs by a single thermalized electronic distribution populating the excited spin states. Hereafter, we prove that this picture is not fully satisfactory. By taking advantage of the large signal/noise ratio of  our time of flight detector \cite{Suppl_mat}, our data show that the measured intensities at energy further above $E_F$ are larger than the calculated ones (two order of magnitude larger), for the same electronic temperature which fits the intensity close to $E_F$.  This \emph{hot} electronic tail calls for a different description.
To address this issue the characteristic relaxation times for both spin states above the Fermi level are compared.



Figure~2~(a) shows the spin resolved intensity difference EDCs ($\Delta$I) measured at 500 fs delay after excitation. The temporal evolution of $\Delta$I is presented in Fig. 2(b) for three energy regions (black arrows in Figure~2~(a)). By fitting a single-exponential decay function to the data, we extract the spin dependent relaxation times. Slightly above $E_F$ (55 meV),  the intensity of the two opposite spins have comparable characteristic relaxation times, $\tau_{red} = 4.6$~ps and $\tau_{blue} = 4.0$~ps. The same behavior is observed for photo-electron kinetic energies up to $255$~meV (second region),  where the two spin populations have the same intensity and $\tau_{red} = \tau_{blue}= 1.8$~ps. On the contrary,  at 370 meV (third region) a faster relaxation for the majority spin component (spin up, in red) is clearly observed, $\tau_{red} = 0.9$~ps and $\tau_{blue} = 2.4$~ps.

Figure~2~(c) displays the spin dependent decay time over the full energy range. In the proximity of  $E_F$ the decay times are comparable \cite{Suppl_mat}.  In the energy region 300-500 meV above $E_F$ the minority spin components persists for longer times. This peculiar behavior at high energy suggests that the out-of-equilibrium electronic properties cannot be simply described in terms of a single thermalized electronic population. In fact, in this simple picture we would expect the same $\tau$ for both spin components along with a monotonic decrease of $\tau$ as a function of the binding energy (see supplementary material \cite{Suppl_mat} and reference \cite{Crepaldi_2012}). 
Hence, a single thermalized electronic population cannot account for the fine structure of $\tau$ that is observed for $E-E_F~>~0.35$ eV: the different spin relaxation times and the local maximum in the blue spin channel. 
In order to reproduce these features we propose to describe the out-of-equilibrium state by a superposition of two distinct thermalized electronic populations.


The photoemission signal at 500 fs, shown in Figure 3 (a), already suggests the existence of two distinct electronic populations. Up to 0.4 eV, the surface contribution associated to the TSS and the SRS dominates, with an effective electronic temperature $T_S(500fs) \sim850$~K. However,  at higher binding  energies the photoemission intensity does not drop according to a 850 K Fermi-Dirac distribution. This is  clearly evidenced by the mismatch between the experimental data and the calculated  spin EDCs represented by the black dashed lines in Fig. 3 (a). The measured spin EDCs display a tail at $E- E_F >$0.6 eV characterized by a small spin un-polarized density of states (DOS) (two orders of magnitude smaller than the surface DOS). This second electronic population has an  effective temperature of $T_B(500fs)=2300$~K. The energy position of this electronic tail suggests that the second electronic subsystem is a \emph{hot} electron gas excited in the high energy side  of the BCB.


The spin- and time-resolved photoemission data further support the presence of two non-interacting electronic systems. We write the photoemission spectrum as the sum of two independent contributions associated to the surface (\emph{S}) and bulk (\emph{B}) :
\begin{equation} \label{eq:twosystems}
\begin{split}
 	I^{PES}_{tot}(\sigma,E,t) &= A_S(\sigma,E) \left|M_S(\sigma,E)\right|^2  f_{FD}(E,T_{S}( t), \mu_S(t) )\\
	& + A_B(\sigma,E) \left|M_B(\sigma,E)\right|^2  f_{FD}(E,T_B(t),0),
\end{split}
\end{equation}
where $T_S\left( t\right)= T_S\left(\infty\right) + \Delta T_S \,\exp\left[ -t/\tau_{TS}\right]$ and $\mu_S\left( t\right)= \mu_S\left(\infty\right) + \Delta \mu_S \,\exp\left[ -t/\tau_{\mu S}\right]$ are the time evolution of the electronic temperature and the chemical potential of the surface states, respectively. For the bulk states, $T_B\left( t\right)= T_B\left(\infty\right) + \Delta T_B \,\exp\left[ -t/\tau_{TB}\right]$.  In good approximation, the chemical potential of the bulk can be kept fixed. The relaxation time constants are left as free parameters.


  \begin{figure}
 \includegraphics[width=0.5\textwidth]{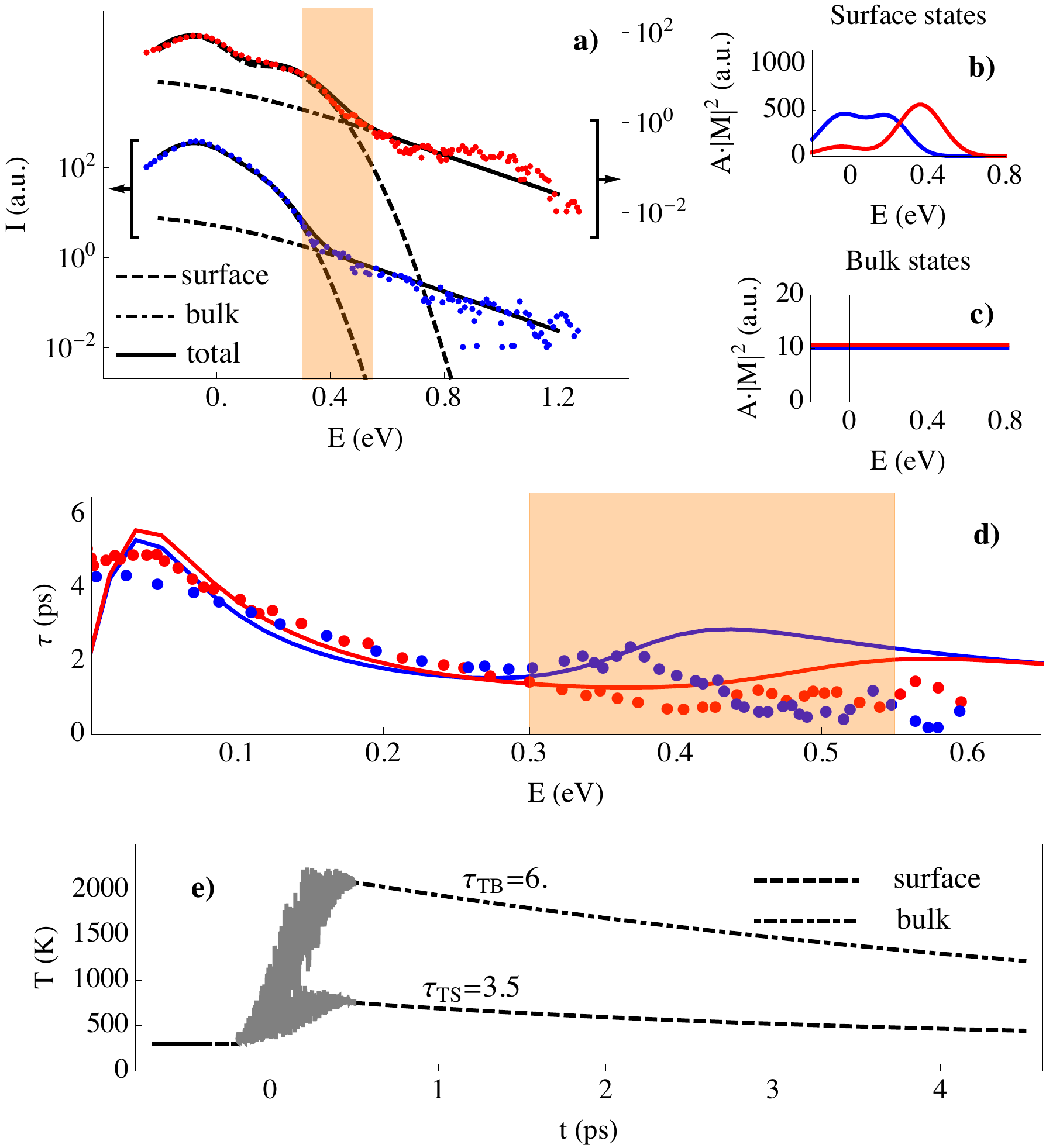}
 \caption{(Colour online)  (a) experimental (points) and theoretical  (full line) spin resolved photoemission intensities, shifted to facilitate the visualization, for the red and blue spin channels. The contributions from the surface and bulks states to the total theoretical photocurrent are highlighted. (b) and (c) optimized  spectral function $A$ multiplied by the photoemission matrix element $\left|M\right|^2$ of the surface states (b) and the bulk states (c) (see supplementary materials for details about the optimization and fit).  (d) experimental (points) and theoretical (full line) energy and spin resolved decay times of the photocurrent intensity. (e) temporal evolution of the electronic temperatures for the surface (dashed line) and bulk states (dashed-dotted line). }
 \label{fig:timezero}
 \end{figure}
 

A fitting procedure is used to extract the energy and spin resolved relaxation times. Figure 3 (b) and (c) show the $A_i(\sigma,E) \left|M_i(\sigma,E)\right|^2$ with $i = S,~B$ functions. The two are optimized starting from the \emph{ab initio} relativistic calculations, in order to better reproduce the experimental spin EDCs after optical excitation  (see supplementary material \cite{Suppl_mat}). 



Figure 3 (d) shows that,  in the energy region close to $E_F$, both the equilibrium and  the time resolved spin ARPES signals are dominated by the surface states contribution, the bulk DOS being two orders of magnitude smaller, while for $E - E_F > 600$ meV  the signal  is dominated by the bulk states, with larger electronic temperature (see supplementary materials  \cite{Suppl_mat}). Interestingly, for $E \sim E_F$ and $E - E_F > 600$ meV, the two spin channels show the same relaxation dynamics, regardless of the different spin polarization of the surface and bulk states. 

For  $300$ meV $ >E - E_F > 600$ meV, on the other hand, highlighted by the orange area in Fig. 3 (d), both subsystems contribute to the total photoemission intensity. The surface state contribution to the blue (red) spin component is smaller (larger) than the bulk one. This results in different spin dynamics. In particular, the local maximum in the characteristic relaxation time of the blue spin component results from the larger relaxation times of the bulk states.  Fig.~3 (d) shows the decay times obtained by the numerical fitting of Eq.~(\ref{eq:twosystems}) where $\mu_S(0)=0.02$~eV, $\tau_{TS}=3.5$~ps, $\tau_{\mu S}=1.7$~ps and $\tau_{TB}=6$~ps were used. Remarkably, the values obtained for the TSS are comparable to those reported in previous TR-ARPES studies \cite{Sobota_PRL_2012, Gedik_PRL_2012, Crepaldi_2012}, whereas the relaxation time of the bulk states is about  twice that  of the surface states. 

Figure 3 (e) clearly shows that the surface (dashed line) and bulk states (dashed-dotted line) behave as independent electronic populations, which thermalize after optical excitation to two different electronic temperatures.  Subsequently, these relax back to equilibrium with  different characteristic times, $\tau_{TS}$ and $\tau_{TB}$. 

In summary, we have investigated the out-of-equilibrium spin and electronic properties of  $\mathrm{Bi_{2}Se_{3}}$. The spin resolution combined with the very high signal/noise level of the time-of-flight spectrometer enables us to reveal novel aspects of the spin and electron dynamics in TIs. We fully map the band structure in the un-occupied density of states and we identify a spin polarized surface resonant state with topological trivial character. Furthermore, the measured weak signal at high energy above the Fermi level cannot be interpreted in terms of a single thermalized electronic population, while it suggests the existence of a second thermalized electronic population. On the basis of \emph{ab initio} photoemission calculations, we attribute the two populations to surface and bulk states respectively, being the latter characterized by a larger effective electronic temperature ($T_B =2300$ K, $T_S = 850$ K). 

Accessing the relaxation dynamics in the time domain, we observe that the two electronic populations have also different characteristic relaxation dynamics. We stress the fact that the different electronic temperature and the different relaxation times indicate that electrons in the surface and bulk states are weakly coupled, and in the investigated temporal window ($\sim$8 ps) they are well described in term of independent populations. Further investigations are required to elucidate the origin of this weak coupling.



This work was supported in part by the Italian Ministry of University and Research under Grant Nos. FIRBRBAP045JF2 and FIRB-RBAP06AWK3 and by the European Community–Research Infrastructure Action under the FP6 "Structuring the European Research Area" Programme through the Integrated Infrastructure Initiative "Integrating Activity on Synchrotron and Free Electron Laser Science" Contract No. RII3-CT-2004-506008. C.C. would like to thanks Phil Rice for his technical support. J.B., H.E. and J.M., members of the COST action MP1306 EUSpec, would like to thank for support by the Deutsche Forschungsgemeinschaft (DFG) within SPP1666, by the BMBF (Project 05K13WMA) and CENTEM (CZ.1.05/2.1.00/03.0088).
C. C and A. C. contributed equally to this work.

\bibliographystyle{prsty}



\begin{thebibliography}{21}
\expandafter\ifx\csname natexlab\endcsname\relax\def\natexlab#1{#1}\fi
\expandafter\ifx\csname bibnamefont\endcsname\relax
  \def\bibnamefont#1{#1}\fi
\expandafter\ifx\csname bibfnamefont\endcsname\relax
  \def\bibfnamefont#1{#1}\fi
\expandafter\ifx\csname citenamefont\endcsname\relax
  \def\citenamefont#1{#1}\fi
\expandafter\ifx\csname url\endcsname\relax
  \def\url#1{\texttt{#1}}\fi
\expandafter\ifx\csname urlprefix\endcsname\relax\def\urlprefix{URL }\fi
\providecommand{\bibinfo}[2]{#2}
\providecommand{\eprint}[2][]{\url{#2}}

\bibitem[{\citenamefont{Hasan and Kane}(2010)}]{Hasan_RMP_2010}
\bibinfo{author}{\bibfnamefont{M.~Z.} \bibnamefont{Hasan}} \bibnamefont{and}
  \bibinfo{author}{\bibfnamefont{C.~L.} \bibnamefont{Kane}},
  \bibinfo{journal}{Rev. Mod. Phys.} \textbf{\bibinfo{volume}{82}},
  \bibinfo{pages}{3046} (\bibinfo{year}{2010}).

\bibitem[{\citenamefont{Qi and Zhang}(2011)}]{Qi_RMP_2011}
\bibinfo{author}{\bibfnamefont{X.-L.} \bibnamefont{Qi}} \bibnamefont{and}
  \bibinfo{author}{\bibfnamefont{S.-C.} \bibnamefont{Zhang}},
  \bibinfo{journal}{Rev. Mod. Phys.} \textbf{\bibinfo{volume}{83}},
  \bibinfo{pages}{1057} (\bibinfo{year}{2011}),
  \urlprefix\url{http://link.aps.org/doi/10.1103/RevModPhys.83.1057}.

\bibitem[{\citenamefont{McIver et~al.}(2012)\citenamefont{McIver, Hsieh,
  Steinberg, Jarillo-Herrero, and Gedik}}]{McIver_Nnano_2012}
\bibinfo{author}{\bibfnamefont{J.~W.} \bibnamefont{McIver}},
  \bibinfo{author}{\bibfnamefont{D.}~\bibnamefont{Hsieh}},
  \bibinfo{author}{\bibfnamefont{H.}~\bibnamefont{Steinberg}},
  \bibinfo{author}{\bibfnamefont{P.}~\bibnamefont{Jarillo-Herrero}},
  \bibnamefont{and} \bibinfo{author}{\bibfnamefont{N.}~\bibnamefont{Gedik}},
  \bibinfo{journal}{Nature Nanotech.} \textbf{\bibinfo{volume}{7}},
  \bibinfo{pages}{96} (\bibinfo{year}{2012}).

\bibitem[{\citenamefont{Sobota et~al.}(2012)\citenamefont{Sobota, Yang,
  Analytis, Chen, Fisher, Kirchmann, and Shen}}]{Sobota_PRL_2012}
\bibinfo{author}{\bibfnamefont{J.~A.} \bibnamefont{Sobota}},
  \bibinfo{author}{\bibfnamefont{S.}~\bibnamefont{Yang}},
  \bibinfo{author}{\bibfnamefont{J.~G.} \bibnamefont{Analytis}},
  \bibinfo{author}{\bibfnamefont{Y.~L.} \bibnamefont{Chen}},
  \bibinfo{author}{\bibfnamefont{I.~R.} \bibnamefont{Fisher}},
  \bibinfo{author}{\bibfnamefont{P.~S.} \bibnamefont{Kirchmann}},
  \bibnamefont{and} \bibinfo{author}{\bibfnamefont{Z.-X.} \bibnamefont{Shen}},
  \bibinfo{journal}{Phys. Rev. Lett.} \textbf{\bibinfo{volume}{108}},
  \bibinfo{pages}{117403} (\bibinfo{year}{2012}).

\bibitem[{\citenamefont{Hajlaoui et~al.}(2012)\citenamefont{Hajlaoui,
  Papalazarou, Mauchain, Lantz, Moisan, Boschetto, Jiang, Miotkowski, Chen,
  Taleb-Ibrahimi et~al.}}]{Perfetti_arxiv_2012}
\bibinfo{author}{\bibfnamefont{M.}~\bibnamefont{Hajlaoui}},
  \bibinfo{author}{\bibfnamefont{E.}~\bibnamefont{Papalazarou}},
  \bibinfo{author}{\bibfnamefont{J.}~\bibnamefont{Mauchain}},
  \bibinfo{author}{\bibfnamefont{G.}~\bibnamefont{Lantz}},
  \bibinfo{author}{\bibfnamefont{N.}~\bibnamefont{Moisan}},
  \bibinfo{author}{\bibfnamefont{D.}~\bibnamefont{Boschetto}},
  \bibinfo{author}{\bibfnamefont{Z.}~\bibnamefont{Jiang}},
  \bibinfo{author}{\bibfnamefont{I.}~\bibnamefont{Miotkowski}},
  \bibinfo{author}{\bibfnamefont{Y.~P.} \bibnamefont{Chen}},
  \bibinfo{author}{\bibfnamefont{A.}~\bibnamefont{Taleb-Ibrahimi}},
  \bibnamefont{et~al.}, \bibinfo{journal}{NanoLett.}
  \textbf{\bibinfo{volume}{12}}, \bibinfo{pages}{3532} (\bibinfo{year}{2012}).

\bibitem[{\citenamefont{Wang et~al.}(2012)\citenamefont{Wang, Hsieh, Sie,
  Steinberg, Gardner, Lee, Jarillo-Herrero, and Gedik}}]{Gedik_PRL_2012}
\bibinfo{author}{\bibfnamefont{Y.~H.} \bibnamefont{Wang}},
  \bibinfo{author}{\bibfnamefont{D.}~\bibnamefont{Hsieh}},
  \bibinfo{author}{\bibfnamefont{E.~J.} \bibnamefont{Sie}},
  \bibinfo{author}{\bibfnamefont{H.}~\bibnamefont{Steinberg}},
  \bibinfo{author}{\bibfnamefont{D.~R.} \bibnamefont{Gardner}},
  \bibinfo{author}{\bibfnamefont{Y.~S.} \bibnamefont{Lee}},
  \bibinfo{author}{\bibfnamefont{P.}~\bibnamefont{Jarillo-Herrero}},
  \bibnamefont{and} \bibinfo{author}{\bibfnamefont{N.}~\bibnamefont{Gedik}},
  \bibinfo{journal}{Phys. Rev. Lett.} \textbf{\bibinfo{volume}{109}},
  \bibinfo{pages}{127401} (\bibinfo{year}{2012}).

\bibitem[{\citenamefont{Crepaldi et~al.}(2012)\citenamefont{Crepaldi, Ressel,
  Cilento, Zacchigna, Grazioli, Berger, Bugnon, Kern, Grioni, and
  Parmigiani}}]{Crepaldi_2012}
\bibinfo{author}{\bibfnamefont{A.}~\bibnamefont{Crepaldi}},
  \bibinfo{author}{\bibfnamefont{B.}~\bibnamefont{Ressel}},
  \bibinfo{author}{\bibfnamefont{F.}~\bibnamefont{Cilento}},
  \bibinfo{author}{\bibfnamefont{M.}~\bibnamefont{Zacchigna}},
  \bibinfo{author}{\bibfnamefont{C.}~\bibnamefont{Grazioli}},
  \bibinfo{author}{\bibfnamefont{H.}~\bibnamefont{Berger}},
  \bibinfo{author}{\bibfnamefont{P.}~\bibnamefont{Bugnon}},
  \bibinfo{author}{\bibfnamefont{K.}~\bibnamefont{Kern}},
  \bibinfo{author}{\bibfnamefont{M.}~\bibnamefont{Grioni}}, \bibnamefont{and}
  \bibinfo{author}{\bibfnamefont{F.}~\bibnamefont{Parmigiani}},
  \bibinfo{journal}{Phys. Rev. B} \textbf{\bibinfo{volume}{86}},
  \bibinfo{pages}{205133} (\bibinfo{year}{2012}),
  \urlprefix\url{http://link.aps.org/doi/10.1103/PhysRevB.86.205133}.

\bibitem[{\citenamefont{Hajlaoui
  et~al.}(2013{\natexlab{a}})\citenamefont{Hajlaoui, Papalazarou, Mauchain,
  Jiang, Miotkowski, Chen, Taleb-Ibrahimi, Perfetti, and
  Marsi}}]{Hajlaoui_EPJ_2013}
\bibinfo{author}{\bibfnamefont{M.}~\bibnamefont{Hajlaoui}},
  \bibinfo{author}{\bibfnamefont{E.}~\bibnamefont{Papalazarou}},
  \bibinfo{author}{\bibfnamefont{J.}~\bibnamefont{Mauchain}},
  \bibinfo{author}{\bibfnamefont{Z.}~\bibnamefont{Jiang}},
  \bibinfo{author}{\bibfnamefont{I.}~\bibnamefont{Miotkowski}},
  \bibinfo{author}{\bibfnamefont{Y.}~\bibnamefont{Chen}},
  \bibinfo{author}{\bibfnamefont{A.}~\bibnamefont{Taleb-Ibrahimi}},
  \bibinfo{author}{\bibfnamefont{L.}~\bibnamefont{Perfetti}}, \bibnamefont{and}
  \bibinfo{author}{\bibfnamefont{M.}~\bibnamefont{Marsi}},
  \bibinfo{journal}{The European Physical Journal Special Topics}
  \textbf{\bibinfo{volume}{222}}, \bibinfo{pages}{1271}
  (\bibinfo{year}{2013}{\natexlab{a}}), ISSN \bibinfo{issn}{1951-6355},
  \urlprefix\url{http://dx.doi.org/10.1140/epjst/e2013-01921-1}.

\bibitem[{\citenamefont{Crepaldi et~al.}(2013)\citenamefont{Crepaldi, Cilento,
  Ressel, Cacho, Johannsen, Zacchigna, Berger, Bugnon, Grazioli, Turcu
  et~al.}}]{Crepaldi_PRB_2013}
\bibinfo{author}{\bibfnamefont{A.}~\bibnamefont{Crepaldi}},
  \bibinfo{author}{\bibfnamefont{F.}~\bibnamefont{Cilento}},
  \bibinfo{author}{\bibfnamefont{B.}~\bibnamefont{Ressel}},
  \bibinfo{author}{\bibfnamefont{C.}~\bibnamefont{Cacho}},
  \bibinfo{author}{\bibfnamefont{J.~C.} \bibnamefont{Johannsen}},
  \bibinfo{author}{\bibfnamefont{M.}~\bibnamefont{Zacchigna}},
  \bibinfo{author}{\bibfnamefont{H.}~\bibnamefont{Berger}},
  \bibinfo{author}{\bibfnamefont{P.}~\bibnamefont{Bugnon}},
  \bibinfo{author}{\bibfnamefont{C.}~\bibnamefont{Grazioli}},
  \bibinfo{author}{\bibfnamefont{I.~C.~E.} \bibnamefont{Turcu}},
  \bibnamefont{et~al.}, \bibinfo{journal}{Phys. Rev. B}
  \textbf{\bibinfo{volume}{88}}, \bibinfo{pages}{121404}
  (\bibinfo{year}{2013}),
  \urlprefix\url{http://link.aps.org/doi/10.1103/PhysRevB.88.121404}.

\bibitem[{\citenamefont{Wang et~al.}(2013)\citenamefont{Wang, Steinberg,
  Jarillo-Herrero, and Gedik}}]{Wang_Science_2013}
\bibinfo{author}{\bibfnamefont{Y.~H.} \bibnamefont{Wang}},
  \bibinfo{author}{\bibfnamefont{H.}~\bibnamefont{Steinberg}},
  \bibinfo{author}{\bibfnamefont{P.}~\bibnamefont{Jarillo-Herrero}},
  \bibnamefont{and} \bibinfo{author}{\bibfnamefont{N.}~\bibnamefont{Gedik}},
  \bibinfo{journal}{Science} \textbf{\bibinfo{volume}{342}},
  \bibinfo{pages}{453} (\bibinfo{year}{2013}),
  \eprint{http://www.sciencemag.org/content/342/6157/453.full.pdf},
  \urlprefix\url{http://www.sciencemag.org/content/342/6157/453.abstract}.

\bibitem[{\citenamefont{Hajlaoui
  et~al.}(2013{\natexlab{b}})\citenamefont{Hajlaoui, Papalazarou, Mauchain,
  Perfetti, Taleb-Ibrahimi, Navarin, Monteverde, Auban-Senzier, Pasquier,
  Moisan et~al.}}]{Marsi_natCom_2014}
\bibinfo{author}{\bibfnamefont{M.}~\bibnamefont{Hajlaoui}},
  \bibinfo{author}{\bibfnamefont{E.}~\bibnamefont{Papalazarou}},
  \bibinfo{author}{\bibfnamefont{J.}~\bibnamefont{Mauchain}},
  \bibinfo{author}{\bibfnamefont{L.}~\bibnamefont{Perfetti}},
  \bibinfo{author}{\bibfnamefont{A.}~\bibnamefont{Taleb-Ibrahimi}},
  \bibinfo{author}{\bibfnamefont{F.}~\bibnamefont{Navarin}},
  \bibinfo{author}{\bibfnamefont{M.}~\bibnamefont{Monteverde}},
  \bibinfo{author}{\bibfnamefont{P.}~\bibnamefont{Auban-Senzier}},
  \bibinfo{author}{\bibfnamefont{C.}~\bibnamefont{Pasquier}},
  \bibinfo{author}{\bibfnamefont{N.}~\bibnamefont{Moisan}},
  \bibnamefont{et~al.}, \bibinfo{journal}{Nat. Comm.}
  \textbf{\bibinfo{volume}{83}}, \bibinfo{pages}{1789}
  (\bibinfo{year}{2013}{\natexlab{b}}).

\bibitem[{\citenamefont{Sobota et~al.}(2014)\citenamefont{Sobota, Yang,
  Leuenberger, Kemper, Analytis, Fisher, Kirchmann, Devereaux, and
  Shen}}]{Sobota2014}
\bibinfo{author}{\bibfnamefont{J.}~\bibnamefont{Sobota}},
  \bibinfo{author}{\bibfnamefont{S.-L.} \bibnamefont{Yang}},
  \bibinfo{author}{\bibfnamefont{D.}~\bibnamefont{Leuenberger}},
  \bibinfo{author}{\bibfnamefont{A.}~\bibnamefont{Kemper}},
  \bibinfo{author}{\bibfnamefont{J.}~\bibnamefont{Analytis}},
  \bibinfo{author}{\bibfnamefont{I.}~\bibnamefont{Fisher}},
  \bibinfo{author}{\bibfnamefont{P.}~\bibnamefont{Kirchmann}},
  \bibinfo{author}{\bibfnamefont{T.}~\bibnamefont{Devereaux}},
  \bibnamefont{and} \bibinfo{author}{\bibfnamefont{Z.-X.} \bibnamefont{Shen}},
  \bibinfo{journal}{Journal of Electron Spectroscopy and Related Phenomena}
  pp.~\bibinfo{pages}{--} (\bibinfo{year}{2014}), ISSN
  \bibinfo{issn}{0368-2048},
  \urlprefix\url{http://www.sciencedirect.com/science/article/pii/S03682048140%
00231}.

\bibitem[{\citenamefont{Niesner et~al.}(2014)\citenamefont{Niesner, Otto,
  Hermann, Fauster, Menshchikova, Eremeev, Aliev, Amiraslanov, Babanly,
  Echenique et~al.}}]{Niesner_PRB_2014}
\bibinfo{author}{\bibfnamefont{D.}~\bibnamefont{Niesner}},
  \bibinfo{author}{\bibfnamefont{S.}~\bibnamefont{Otto}},
  \bibinfo{author}{\bibfnamefont{V.}~\bibnamefont{Hermann}},
  \bibinfo{author}{\bibfnamefont{T.}~\bibnamefont{Fauster}},
  \bibinfo{author}{\bibfnamefont{T.~V.} \bibnamefont{Menshchikova}},
  \bibinfo{author}{\bibfnamefont{S.~V.} \bibnamefont{Eremeev}},
  \bibinfo{author}{\bibfnamefont{Z.~S.} \bibnamefont{Aliev}},
  \bibinfo{author}{\bibfnamefont{I.~R.} \bibnamefont{Amiraslanov}},
  \bibinfo{author}{\bibfnamefont{M.~B.} \bibnamefont{Babanly}},
  \bibinfo{author}{\bibfnamefont{P.~M.} \bibnamefont{Echenique}},
  \bibnamefont{et~al.}, \bibinfo{journal}{Phys. Rev. B}
  \textbf{\bibinfo{volume}{89}}, \bibinfo{pages}{081404}
  (\bibinfo{year}{2014}),
  \urlprefix\url{http://link.aps.org/doi/10.1103/PhysRevB.89.081404}.

\bibitem[{\citenamefont{Hsieh et~al.}(2009{\natexlab{a}})\citenamefont{Hsieh,
  Xia, Qian, Wray, Dil, Meier, Osterwalder, Patthey, Checkelsky, Ong
  et~al.}}]{Hsieh_nature_2009}
\bibinfo{author}{\bibfnamefont{D.}~\bibnamefont{Hsieh}},
  \bibinfo{author}{\bibfnamefont{Y.}~\bibnamefont{Xia}},
  \bibinfo{author}{\bibfnamefont{D.}~\bibnamefont{Qian}},
  \bibinfo{author}{\bibfnamefont{L.}~\bibnamefont{Wray}},
  \bibinfo{author}{\bibfnamefont{J.~H.} \bibnamefont{Dil}},
  \bibinfo{author}{\bibfnamefont{F.}~\bibnamefont{Meier}},
  \bibinfo{author}{\bibfnamefont{J.}~\bibnamefont{Osterwalder}},
  \bibinfo{author}{\bibfnamefont{L.}~\bibnamefont{Patthey}},
  \bibinfo{author}{\bibfnamefont{J.~G.} \bibnamefont{Checkelsky}},
  \bibinfo{author}{\bibfnamefont{N.~P.} \bibnamefont{Ong}},
  \bibnamefont{et~al.}, \bibinfo{journal}{Nature}
  \textbf{\bibinfo{volume}{460}}, \bibinfo{pages}{1101}
  (\bibinfo{year}{2009}{\natexlab{a}}).

\bibitem[{\citenamefont{Hsieh et~al.}(2009{\natexlab{b}})\citenamefont{Hsieh,
  Xia, Wray, Qian, Pal, Dil, Osterwalder, Meier, Bihlmayer, Kane
  et~al.}}]{Hsieh_Science_2009}
\bibinfo{author}{\bibfnamefont{D.}~\bibnamefont{Hsieh}},
  \bibinfo{author}{\bibfnamefont{Y.}~\bibnamefont{Xia}},
  \bibinfo{author}{\bibfnamefont{L.}~\bibnamefont{Wray}},
  \bibinfo{author}{\bibfnamefont{D.}~\bibnamefont{Qian}},
  \bibinfo{author}{\bibfnamefont{A.}~\bibnamefont{Pal}},
  \bibinfo{author}{\bibfnamefont{J.~H.} \bibnamefont{Dil}},
  \bibinfo{author}{\bibfnamefont{J.}~\bibnamefont{Osterwalder}},
  \bibinfo{author}{\bibfnamefont{F.}~\bibnamefont{Meier}},
  \bibinfo{author}{\bibfnamefont{G.}~\bibnamefont{Bihlmayer}},
  \bibinfo{author}{\bibfnamefont{C.~L.} \bibnamefont{Kane}},
  \bibnamefont{et~al.}, \bibinfo{journal}{Science}
  \textbf{\bibinfo{volume}{323}}, \bibinfo{pages}{919}
  (\bibinfo{year}{2009}{\natexlab{b}}).

\bibitem[{Sup(2014)}]{Suppl_mat}
\emph{\bibinfo{title}{See supplementary information at url XXX}}
  (\bibinfo{year}{2014}).

\bibitem[{\citenamefont{Jozwiak et~al.}(2013)\citenamefont{Jozwiak, Park,
  Gotlieb, Hwang, Lee, Louie, Denlinger, Rotundu, Birgeneau, Hussain
  et~al.}}]{Jozwiak_NatPhys_2013}
\bibinfo{author}{\bibfnamefont{C.}~\bibnamefont{Jozwiak}},
  \bibinfo{author}{\bibfnamefont{C.-H.} \bibnamefont{Park}},
  \bibinfo{author}{\bibfnamefont{K.}~\bibnamefont{Gotlieb}},
  \bibinfo{author}{\bibfnamefont{C.}~\bibnamefont{Hwang}},
  \bibinfo{author}{\bibfnamefont{D.-H.} \bibnamefont{Lee}},
  \bibinfo{author}{\bibfnamefont{S.~G.} \bibnamefont{Louie}},
  \bibinfo{author}{\bibfnamefont{J.~D.} \bibnamefont{Denlinger}},
  \bibinfo{author}{\bibfnamefont{C.~R.} \bibnamefont{Rotundu}},
  \bibinfo{author}{\bibfnamefont{R.~J.} \bibnamefont{Birgeneau}},
  \bibinfo{author}{\bibfnamefont{Z.}~\bibnamefont{Hussain}},
  \bibnamefont{et~al.}, \bibinfo{journal}{Nat. Phys.} p.
  \bibinfo{pages}{293–298} (\bibinfo{year}{2013}).

\bibitem[{\citenamefont{Zhu et~al.}(2014)\citenamefont{Zhu, Veenstra,
  Zhdanovich, Schneider, Okuda, Miyamoto, Zhu, Namatame, Taniguchi, Haverkort
  et~al.}}]{Zhu_PRL_2014}
\bibinfo{author}{\bibfnamefont{Z.-H.} \bibnamefont{Zhu}},
  \bibinfo{author}{\bibfnamefont{C.}~\bibnamefont{Veenstra}},
  \bibinfo{author}{\bibfnamefont{S.}~\bibnamefont{Zhdanovich}},
  \bibinfo{author}{\bibfnamefont{M.~P.} \bibnamefont{Schneider}},
  \bibinfo{author}{\bibfnamefont{T.}~\bibnamefont{Okuda}},
  \bibinfo{author}{\bibfnamefont{K.}~\bibnamefont{Miyamoto}},
  \bibinfo{author}{\bibfnamefont{S.-Y.} \bibnamefont{Zhu}},
  \bibinfo{author}{\bibfnamefont{H.}~\bibnamefont{Namatame}},
  \bibinfo{author}{\bibfnamefont{M.}~\bibnamefont{Taniguchi}},
  \bibinfo{author}{\bibfnamefont{M.~W.} \bibnamefont{Haverkort}},
  \bibnamefont{et~al.}, \bibinfo{journal}{Phys. Rev. Lett.}
  \textbf{\bibinfo{volume}{112}}, \bibinfo{pages}{076802}
  (\bibinfo{year}{2014}),
  \urlprefix\url{http://link.aps.org/doi/10.1103/PhysRevLett.112.076802}.

\bibitem[{\citenamefont{Braun}(1996)}]{Bra96}
\bibinfo{author}{\bibfnamefont{J.}~\bibnamefont{Braun}}, \bibinfo{journal}{Rep.
  Prog. Phys.} \textbf{\bibinfo{volume}{59}}, \bibinfo{pages}{1267}
  (\bibinfo{year}{1996}).

\bibitem[{\citenamefont{Pendry}(1974)}]{Pen74}
\bibinfo{author}{\bibfnamefont{J.~B.} \bibnamefont{Pendry}},
  \emph{\bibinfo{title}{Low Energy Electron Diffraction}}
  (\bibinfo{publisher}{Academic Press}, \bibinfo{year}{1974}).

\bibitem[{\citenamefont{S\'anchez-Barriga
  et~al.}(2014)\citenamefont{S\'anchez-Barriga, Varykhalov, Braun, Xu,
  Alidoust, Kornilov, Min\'ar, Hummer, Springholz, Bauer
  et~al.}}]{Minar_PRX_2014}
\bibinfo{author}{\bibfnamefont{J.}~\bibnamefont{S\'anchez-Barriga}},
  \bibinfo{author}{\bibfnamefont{A.}~\bibnamefont{Varykhalov}},
  \bibinfo{author}{\bibfnamefont{J.}~\bibnamefont{Braun}},
  \bibinfo{author}{\bibfnamefont{S.-Y.} \bibnamefont{Xu}},
  \bibinfo{author}{\bibfnamefont{N.}~\bibnamefont{Alidoust}},
  \bibinfo{author}{\bibfnamefont{O.}~\bibnamefont{Kornilov}},
  \bibinfo{author}{\bibfnamefont{J.}~\bibnamefont{Min\'ar}},
  \bibinfo{author}{\bibfnamefont{K.}~\bibnamefont{Hummer}},
  \bibinfo{author}{\bibfnamefont{G.}~\bibnamefont{Springholz}},
  \bibinfo{author}{\bibfnamefont{G.}~\bibnamefont{Bauer}},
  \bibnamefont{et~al.}, \bibinfo{journal}{Phys. Rev. X}
  \textbf{\bibinfo{volume}{4}}, \bibinfo{pages}{011046} (\bibinfo{year}{2014}),
  \urlprefix\url{http://link.aps.org/doi/10.1103/PhysRevX.4.011046}.

\end{thebibliography}


\end{document}